\documentclass[]{interact}

\usepackage{epstopdf}
\usepackage[caption=false]{subfig}

\usepackage[numbers,sort&compress,merge]{natbib}
\bibpunct[, ]{[}{]}{,}{n}{,}{,}
\renewcommand\bibfont{\fontsize{10}{12}\selectfont}

\theoremstyle{plain}

\theoremstyle{definition}

\theoremstyle{remark}

\usepackage[numbers,sort&compress,merge]{natbib}
\bibpunct[, ]{[}{]}{,}{n}{,}{,}
\renewcommand\bibfont{\fontsize{10}{12}\selectfont}

\begin{document}


\title{
  Vibrational effects in the quantum dynamics of the
  H + D$_2^+$ charge transfer reaction
}

\author{
  \name{
    O. Roncero\textsuperscript{a}\thanks{CONTACT O. Roncero. Email: octavio.roncero@csic.es}
       , V. Andrianarijaona\textsuperscript{b,c}
    , A. Aguado\textsuperscript{d} and C. Sanz-Sanz\textsuperscript{d}
    }
    \affil{\textsuperscript{a}Instituto de F{\'\i}sica Fundamental, IFF-CSIC, c/ Serrano 123, 28006 Madrid, Spain,
      \textsuperscript{b} Department of Physics , Pacific Union College, Angwin, California, United States,
      \textsuperscript{c} Current address: 
      Department of Physics and Engineering, Southern Adventist University, Collegedale, Tennessee, United States
      \textsuperscript{d}Unidad Asociada UAM-CSIC,
                       Departamento de Qu{\'\i}mica F{\'\i}sica Aplicada, Facultad de
                      Ciencias M-14, Universidad Aut\'onoma de Madrid, 28049, Madrid, Spain 
    }
}

\maketitle

This is an accepted manuscript of an article that will be published by Taylor \& Francis in ``Molecular Physics''
on 2021, will be avilable on line: http://wwww.tandfonline.com/10.1080/00268976.2021.1948125

\vspace*{0.5cm}

\begin{abstract}
  The H + D$_2^+$(v=0,1 and 2) charge transfer reaction is studied using an accurate wave packet method,
  using  recently proposed coupled  diabatic potential energy surfaces.
  The state-to-state cross section is obtained for three 
   different channels:  non-reactive charge transfer, reactive charge transfer,
  and exchange reaction.
  The three processes
  proceed via the electronic transition from the first excited to the ground electronic state.
  The cross section for the three processes increases with the 
  initial vibrational excitation.
  The non-reactive charge transfer process is the dominant channel, whose branching ratio increases
  with collision energy, and it compares well
  with experimental measurements at collision energies around 0.5 eV.
  For lower energies the experimental cross section
  is considerably higher, suggesting that it corresponds to higher vibrational excitation of D$_2^+$(v)
  reactants. Further experimental studies of this reaction and isotopic variants are needed, where conditions are
  controlled to obtain a better analysis of the vibrational  effects of the D$_2^+$ reagents. 
\end{abstract}

\begin{keywords}
charge transfer, quantum dynamics, non-adiabatic dynamics
\end{keywords}

\section{Introduction}

H$_2$ is the most abundant molecule in the interstellar medium.
The gas phase routes to form molecular hydrogen present very slow rate constants,
and its formation in local galaxies is
attributed to reactions on cosmic grains and ices\cite{Glover:03,Wakelam-etal:17}.
However, in environments where grains and ice do not exist,
like in the Early Universe,
one key process is the formation of H$_2$ in gas phase. One of such processes is the
charge transfer reaction 
\begin{eqnarray}\label{formation-H2}
  {\rm H} + {\rm H}_2^+ \rightarrow {\rm H}_2 + {\rm H}^+.
\end{eqnarray}
This reaction has been studied experimentally  at rather high
energies, $>$  1 keV, by
Karpas {\it et al}\cite{Karpas-etal:79} and
by McCartney {\it et al.}\cite{McCartney-etal:99}. 
Andrianarijaona {\it et al}\cite{Andrianarijaona-etal:09,Andrianarijaona-etal:19}
measured the  H + D$_2^+$ $\rightarrow$ H$^+$ +D$_2$
charge transfer cross section in a  broad energy range, from  0.1eV/u-10 keV/u,
thus including energies more relevant for the astrophysical environment.
At this point it is worth mentioning that the D$_2^+$ reactant  is excited vibrationally
since it is generated 
by photoionization or electronic impact. The main goal of this work is
to compare the simulated cross sections
for each  D$_2^+$(v) vibrational state with the experimental measurements.
In this line, the measurements
made in the  H + D$_2^+$ reaction \cite{Andrianarijaona-etal:09,Andrianarijaona-etal:19}
can then be considered as benchmark for the comparison with the theoretical simulations,
which conclusions can be generalized to other isotopes and to similar reactions
for other cations.

From the theoretical point of view, reaction~(\ref{formation-H2}) has been
studied by several authors
\cite{Last-etal:97,Kamisaka-etal:02,Krstic:02,Krstic-Janev:03,Errea-etal:05,Ghosh-etal:21,Sanz-Sanz-etal:21},
{
  using different and increasingly more accurate coupled diabatic potential energy surfaces
  (PESs)\cite{Preston-Tully:71,Ichihara:95,Ushakov-etal:01,Mukherjee-etal:14,Mukherjee-etal:19}.
  }
Last {\it et al.} \cite{Last-etal:97} 
  used  a quantum method based on negative imaginary potential combined
  with a variational quantum method in a L$^2$
  basis set, with the helicity decoupling approximation,
  and using  approximated PESs.
  Krstic\cite{Krstic:02} used a close coupling method
  based on the infinite order sudden approximation (IOSA) with Delves hyperspherical
  coordinates. Gosh {\it et al.} \cite{Ghosh-etal:21}
  and Sanz-Sanz {\it et al.} \cite{Sanz-Sanz-etal:21}
  used accurate quantum wave packet methods, in hyperspherical and Jacobi coordinates respectively,
  and used different coupled diabatic PESs based on accurate {\it ab initio} calculations.
  All these calculations showed an important enhancement on a particular vibrational state
  of H$_2$ due to a quasi-degeneracy of H$_2^+$(v=0) and H$_2$(v'=4). This effect
  depends strongly on the long-range behavior of the PESs,
  very accurately described in the work of Aguado {\it et al}\cite{Aguado-etal:21}, and already
  used by Sanz-Sanz {\it et al.} \cite{Sanz-Sanz-etal:21}.
  In this latter work, the charge transfer reaction was studied for  several isotopic variants.
  Among them
  the H+ D$_2^+$(v=0) $\rightarrow$ H$^+$ + D$_2$ charge transfer reaction was studied
  and compared with the experimental values \cite{Andrianarijaona-etal:09,Andrianarijaona-etal:19},
  showing  good agreement for collision energies around 0.5 eV, but some differences appear at
  lower energies. These differences were attributed to vibrational excitation of the D$_2^+$, not
  taken into account by Sanz-Sanz {\it et al.} \cite{Sanz-Sanz-etal:21}. The goal of this
  work is to analyze the effect of vibrational excitation of D$_2^+$, which is expected to present
  important discrepancies due to the variation of the energy difference between the initial D$_2^+$(v)
  and final D$_2$(v') vibrational levels.

  The manuscript is distributed as follows. A  brief description of the theoretical
  method is presented in section 2. In section 3, the quantum results are described and compared
  with the experimental results. Finally, in section 4 are outlined some conclusions.

  \section{Theoretical method}

\begin{figure}
\centering
\subfloat[Potential of D$_2$ and D$_2^+$  at very long distance of H atom, showing the vibrational levels on each.
The energies of D$^+_2$(v) are 2.056, 2.251 and 2.439 eV, for v=0, 1 and 2, respectively.
The energies of D$_2$(v') are 1.902, 2.202, 2.487 and 2.758 for v'= 5,6, 7 and 8, respectively.]{%
\resizebox*{14cm}{!}{\includegraphics{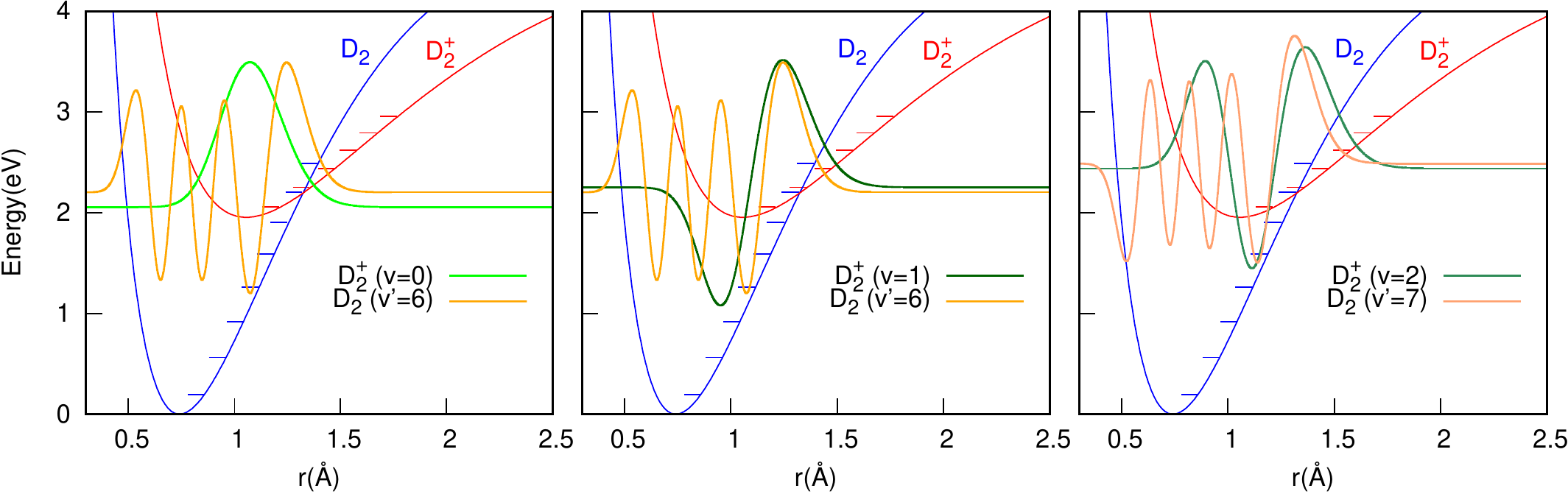}}}\hspace{5pt}
\subfloat[Adiabatic potential energies of the first 2 singlet electronic states as a function
of R, for $r=r_c$ and at a T-shaped configuration. The diabatic coupling between the two diabatic energies
are also shown.]{%
\resizebox*{7cm}{!}{\includegraphics{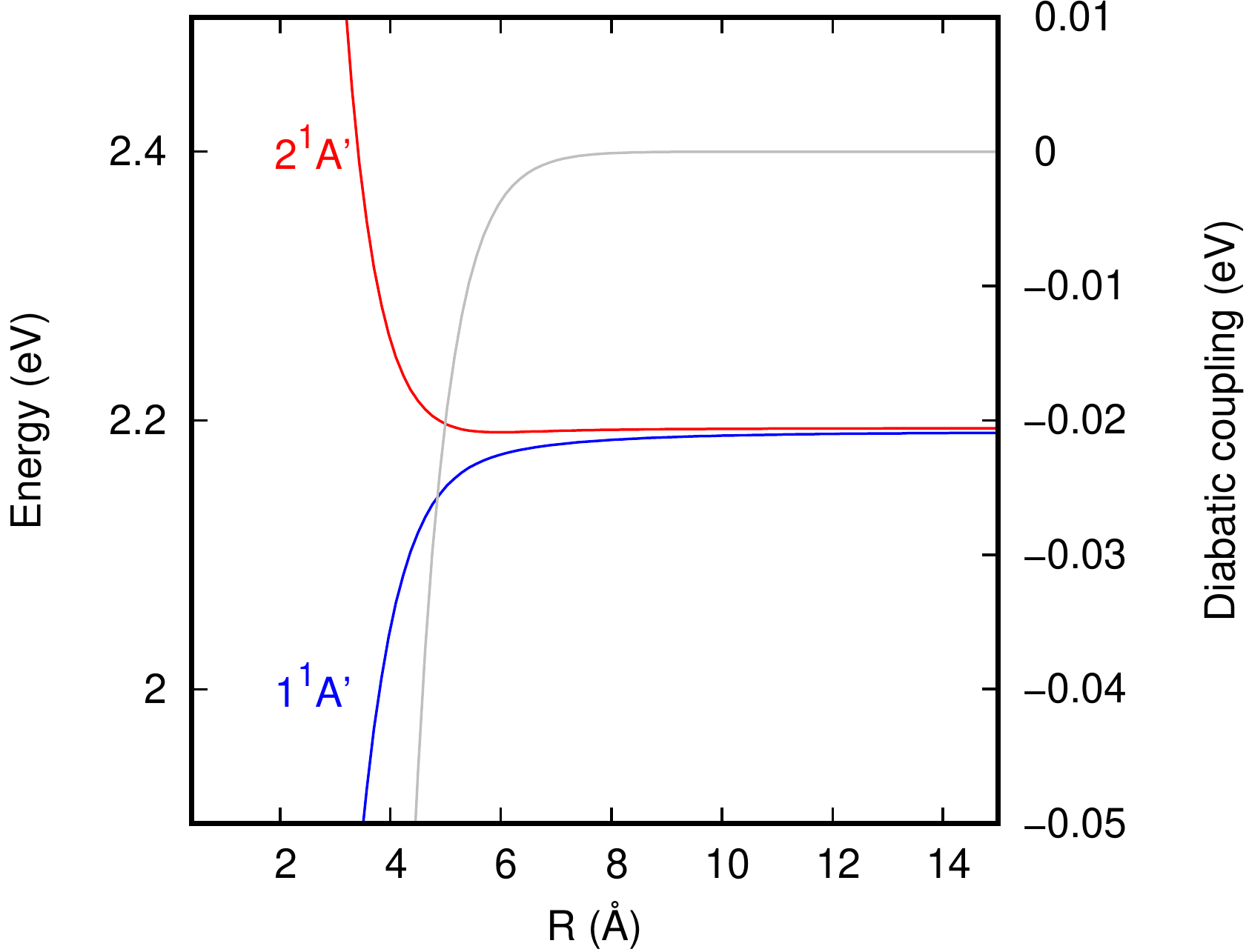}}}
\caption{Features of the 3x3 PES's describing the H+D$_2^+ $ $\rightarrow$ H$^+$ + D$_2$
  charge transfer process, where  $r$ is the D$_2$ internuclear distance and $R$ is the distance
  between H and the center-of-mass of D$_2$.
 } \label{pes-figure}
\end{figure}

  In this work we use the coupled diabatic PESs developed by Aguado {\it et al.} \cite{Aguado-etal:21}
  based on a Diatomics-in-Molecules (DIM) 3$\times$3 matrix description\cite{Ellison-etal:63,Tully:80},
  corrected with diagonal and non-diagonal
   three body-terms \cite{Viegas2007:074309}
   to fit the energies obtained in Multi-Reference Configuration interaction (MRCI) calculations with complete
   basis set extrapolation (CBS).
   { Dealing with a system of two electrons, MRCI calculations give full configuration interactions (FCI)
     results within the space spanned by the orbital set. To obtain near FCI results cfor complete basis set, a complete
     basis set extrapolation method was performed.}
   These PESs include the long-range interaction very accurately.
  The dominant terms for reactants H + H$_2^+$  channel   are the charge-induced dipole
  and charge-induced quadrupole dispersion interactions, varying  as $R^{-4}$ and $R^{-6}$ respectively.
 For the products H$^+$ + H$_2$  channel, the main long-range terms are
  the charge-quadrupole and the charge-induced dipole dispersion energies,
  which vary  as $R^{-3}$ and $R^{-4}$ respectively.
  The DIM diabatic representation is non diagonal for the description of any  H$_2^+$ + H fragment.
  Therefore, a transformation to a new diabatic representation should be done,
  in which the PESs are diagonal in the reactants channel while they are
  non-diagonal in the two product rearrangement channels,
  as described by Sanz-Sanz {\it et al.} \cite{Sanz-Sanz-etal:21}.
  The features of the potential in the reactant channel,
  as a function of Jacobi r variable, are shown in Fig.~\ref{pes-figure}.
   At long distances, the potentials of  D$_2$ and D$_2^+$ cross at $r=r_c$=1.323\AA,
  as shown in Fig.~\ref{pes-figure}.a.
  The  amplitude of D$_2^+(v)$  vibrational state is non-zero at this distance,
  and, in this outer classical turning point region,
  the D$_2^+(v)$ levels have a matching overlap with different D$_2$(v') vibrational
  states. As the two reactants approach each other,  
  the degeneracy between the two first adiabatic states dissappears:
  the ground state gets stabilized forming the
  very stable HD$_2^+$ system, which correlates to H$^+$+D$_2$ and D$^+$+HD
  adiabatic asymptotes, while the
  excited adiabatic state becomes repulsive (see Fig.~\ref{pes-figure}.b).

  The reactive and charge transfer collisions have been studied with a quantum wave packet method
  using the MADWAVE3 code \cite{Zanchet-etal:09b,MADWAVE3:21}, using the parameters listed in Table 1 of
  Ref.~\cite{Sanz-Sanz-etal:21}. The wave packet is propagated for each total angular momentum $J$,
  using a modified Chebyshev propagator\cite{Mandelshtam-Taylor:95,Kroes-Neuhauser:96,Chen-Guo:96,Gray-Balint-Kurti:98,Gonzalez-Lezana-etal:05}.
  The wave packet is represented in reactant Jacobi coordinates
  and is transformed to product Jacobi coordinates \cite{Gomez-Carrasco-Roncero:06} at each iteration
  to extract  the individual state-to-state reaction
  probabilities, $P^{J}_{\alpha vj\Omega\rightarrow \alpha' v'j'\Omega'}(E)$,
  where $\alpha$ denotes the rearrangement channel and the electronic state, $v,j$ correspond to the
  vibrational and rotational state of the diatomic fragment, and
  $\Omega$ is the projection
  of the total angular momentum in the z-axis of the body-fixed frame. The integral state-to-state
  reactive and inelastic cross sections are calculated using the partial wave expansion as
 \begin{eqnarray}
   \sigma_{\alpha v j\rightarrow \alpha' v' j'}(E)&=& q_e
{\pi  k_{\alpha vj}^{-2}\over {(2j+1)}}\sum_{J\Omega\Omega'} (2J+1)
 P^{J}_{\alpha vj\Omega\rightarrow \alpha' v'j'\Omega'}(E)  ,  \nonumber\\
&&
 \end{eqnarray}
 where $q_e=1/4$ is the electronic partition function - note that this factor was not
 included in the previous study \cite{Sanz-Sanz-etal:21} -. There are four  electronic
 states correlating  to H($^2$S)+D$_2^+(^2\Sigma_g^+)$ asymptote, one singlet and three triplet states.
 The three triplet states  are not connected to the H$^+$ + H$_2$(X$^1\Sigma_g^+$) channel, and therefore
 cannot undergo charge transfer reactions.
 In this study, $P^{J}(E)$ are calculated with the MADWAVE3 code
 for all values with $J<15$ 
 and  for  $J$ = 20, 30, 40, 50, 60, 70 and 80. The rest of reaction probabilities
 for all the intermediate $J's$ are obtained by a linear interpolation
 based on the J-shifting approximation, as described before \cite{Aslan-etal:12,Zanchet-etal:13},
 to save computational effort. In these calculations, the
 maximum projection, $\Omega$, considered in the dynamic calculations is 23.
   { 
     In a previous study\cite{Sanz-Sanz-etal:21},  the role of the trunction of $\Omega_{max}$ was checked,
     giving excellent results
  for the NCRT channel, and an accuracy better than 2-3 $\%$ for the other two channels.
  }

 In what follows, three  processes are distinguished
\begin{eqnarray}\label{processes}
   \begin{array}{lccr}
     {\rm H} + {\rm D}_2^+(v,j=0,J)& \longrightarrow &  {\rm H}^+ + {\rm D}_2(v',j')&  {\tiny {\rm non-reactive\, charge\, transfer (NRCT)}}\\
     &&&\\
                      & \longrightarrow &  {\rm HD}^+(v',j') + {\rm D} & {\tiny {\rm exchange\, reaction  (ER)}}\\
     &&&\\
                      & \longrightarrow & {\rm HD}(v',j') + {\rm D}^+ & {\tiny {\rm reactive\, charge\, transfer (RCT)}}.\\
   \end{array}
\end{eqnarray}
Charged diatomic products correspond to the first excited electronic singlet adiabatic state, 2$^1A'$,
while the product with a charged atomic fragment corresponds to the ground adiabatic electronic singlet state, 1$^1A'$.

The H($^2S$)+ H$_2^+(^2\Sigma_g)$ entrance channel also correlates with the lowest triplet state of H$_3^+$($^3A'$). This triplet
state presents a shallow well at collinear geometry, and has access to the exchange products, as described
before\cite{Aguado-etal:21,Sanz-etal:01}. Therefore,
the triplet state contributes to the exchange reaction, ER, but does not contribute to the other two charge transfer
processes, NRCT and RCT.
  { In the present calculations dynamical calculations are all performed for the singlet states.
    Therefore, the cross section for the RCT channel presented below is not complete, since they
    should include the contributions arising from the triplet states.
  }

\section{Results}

\begin{figure}
\centering
{%
  {\includegraphics[width=0.95\linewidth]{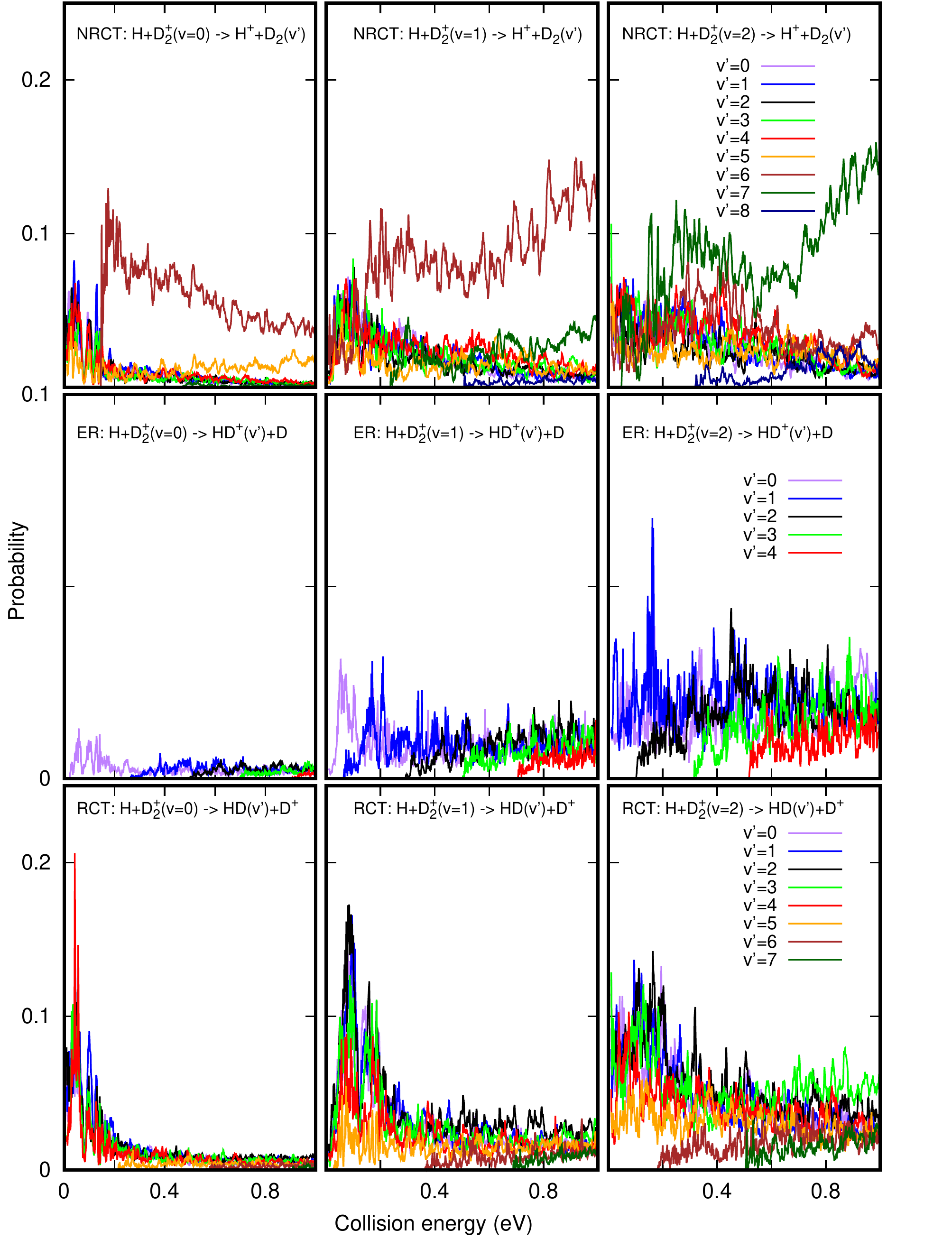}}}
\caption{Vibrationally resolved reaction probability for $J$=0  H+D$_2^+$(v=0,j=0) (left panels), H+D$_2^+$(v=1,j=0) (middle panels),
    and H +D$_2^+$(v=2) (right panels), 
    towards the inelastic charge transfer, NRCT (H$^+$+D$_2$, in the top panels),
    the exchange reactive channel, ER (H+HD$^+$, in the middle panels) and reactive charge transfer, RCT, channels
   (D$^+$ +HD, in the bottom panels).} \label{probJ0}
\end{figure}

The vibrationally resolved reaction probabilities for $J$=0 are shown in Fig.~\ref{probJ0}
for H+D$_2^+$(v) collisions, for v=0,1 and 2,  in the left, middle and right panels, respectively.
The probabilities for the individual final vibrational states, v',  are included
for the three different processes, NRCT (top panels), ER (middle panels)
and RCT (bottom panels). The reaction in the excited adiabatic electronic state
has an extremely
high barrier. However, as the two fragments get closer, the electronic
couplings bring a large portion of the wave packet to the ground adiabatic state correlating
to the two charge transfer channels, NRCT and RCT.
The  NRCT  reaction probabilities
are very close to those of  RCT at low energies,
except when the D$_2$(v') level is close to the D$_2^+$(v) state considered, D$_2$(v'=6) for v=0 and 1 and D$_2$(v'=7) for v=2.
According to  Fig.~\ref{pes-figure}.a,  the final population
of D$_2$(v') with an energy slightly above than that of the initial D$_2^+$(v)
is always enhanced.
This is due to the interaction with the H atom,
which  stabilizes 
the ground electronic state, connected to H$^+$+ D$_2$ asymptote, while
the excited state becomes repulsive (see Fig.~\ref{pes-figure}.b).

For lower energies, the amplitude transfered to the ground electronic state enters the deep
well of HD$_2^+$, forming long lived resonances. In this well, the energy is transfered
among all possible modes and the reaction becomes nearly statistical, as reported
for the reaction dynamics in the ground electronic state
\cite{Gonzalez-Lezana-etal:05,Gonzalez-Lezana-etal:06,Gonzalez-Lezana-Honvault:14}.
This explains why NRCT and RCT show similar
populations, except for  near-resonant v' levels.
This also allows the transition back to the excited electronic
state and explains the appearance of the ER probabilities in the middle panels
of Fig.~\ref{probJ0}, that cannot 
reach directly the excited adiabatic electronic state.

The amplitude of the D$_2^+$(v) vibrational state extends over larger
radial distances as v increases, becoming larger at the region of the crossing between
the two electronic states. This can explain why all the probabilities, NRCT, ER and RCT,
increase with v.

\begin{figure}
\centering
{%
  {\includegraphics[width=0.95\linewidth]{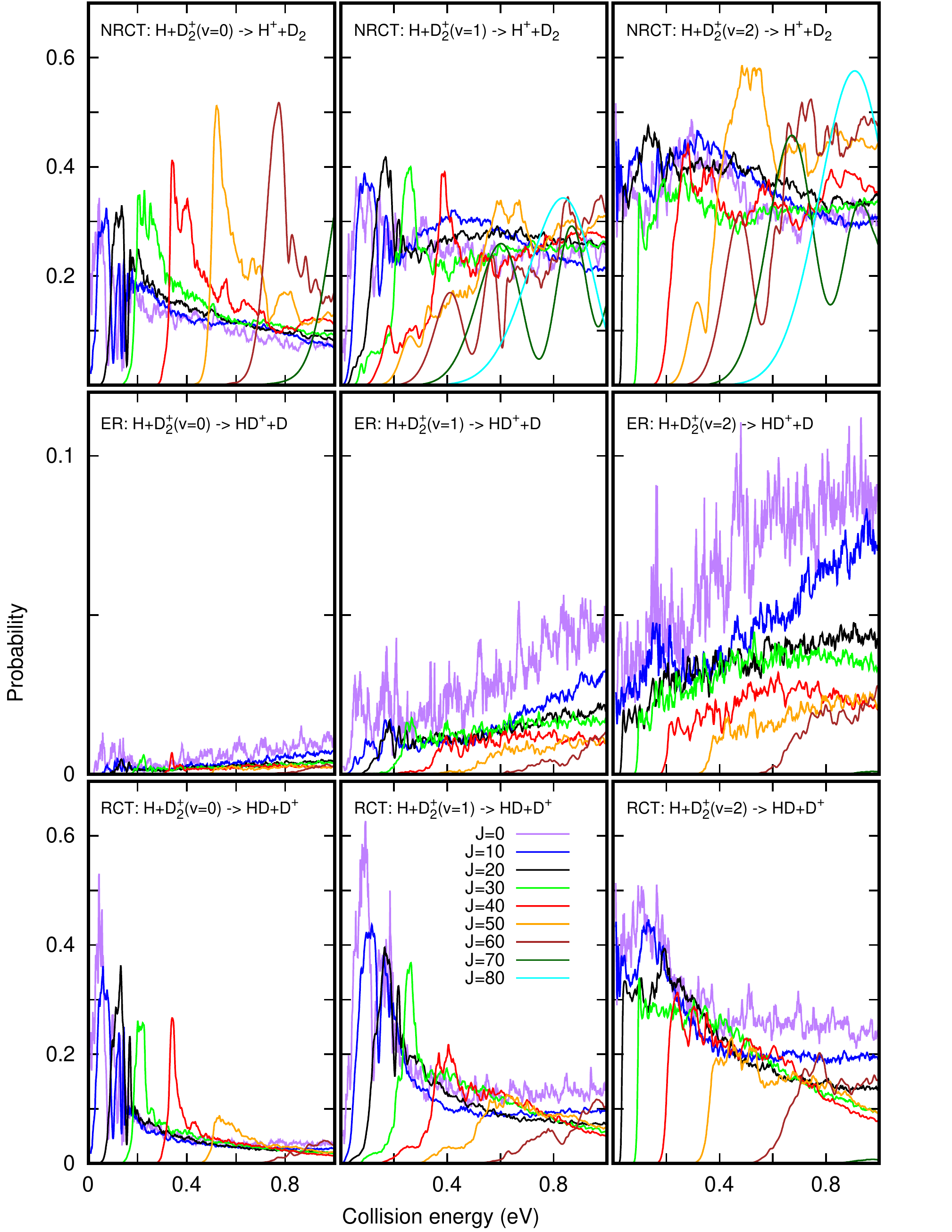}}}
\caption{Reaction probability for different $J$ values  in H+D$_2^+$(v=0) (left panels), H+D$_2^+$(v=1) (middle panels),
    and H +D$_2^+$(v=2) (right panels), 
    towards the inelastic charge transfer, NRCT (H$^+$+D$_2$, in the top panels),
    the exchange reactive channel, ER (H+HD$^+$, in the middle panels) and the reactive charge transfer, RCT, channels
   (D$^+$ +HD, in the bottom panels).} \label{probJ}
\end{figure}

As the total angular momentum $J$ increases, the situation gradually changes.
 The reaction probabilites for the three vibrational
states studied are shown in Fig.~\ref{probJ} for the 3 different rearrangent channels.
As the rotational barrier increases, the reaction probabilities shift towards
higher energy, as expected. This shift, however, depends on the mechanism. Thus,
the two reactive processes, ER and RCT, have always larger shifts or effective barriers.
As the barrier increases with $J$, the reaction probabilities  decreases.

This situation is somehow different for the NRCT channel, which is essentially inelastic
and has lower effective barriers. As shown in the top panel of Fig.~\ref{probJ},
the effective barrier decreases progressively when going from v=0 to v=2. The reason
is that the energy difference between the initial (v) and final (v') vibrational levels also
decreases. Thus, the electronic coupling becomes more effective at longer distances for higher v',
because lower couplings can induce the transition.

The resonant structure appearing at low energy progressively dissapears as 
$J$ increases because the rotational barrier increases
and the wave packet can not reach the deep well in the ground electronic state.
Finally, as the reactive processes decrease, the NRCT proccess increases in intensity with $J$, becoming
dominant.
{ The lower effective barriers observed for the NCRT channel makes necessary to include more partial waves $J$
    to converge the corresponding cross section in Eq.~(2), up to $J$=140, and they were extrapolated
    using the J-shifting
    approximation from the calculated $J$=80.
 }

\begin{figure}
\centering
{%
  {\includegraphics[width=0.75\linewidth]{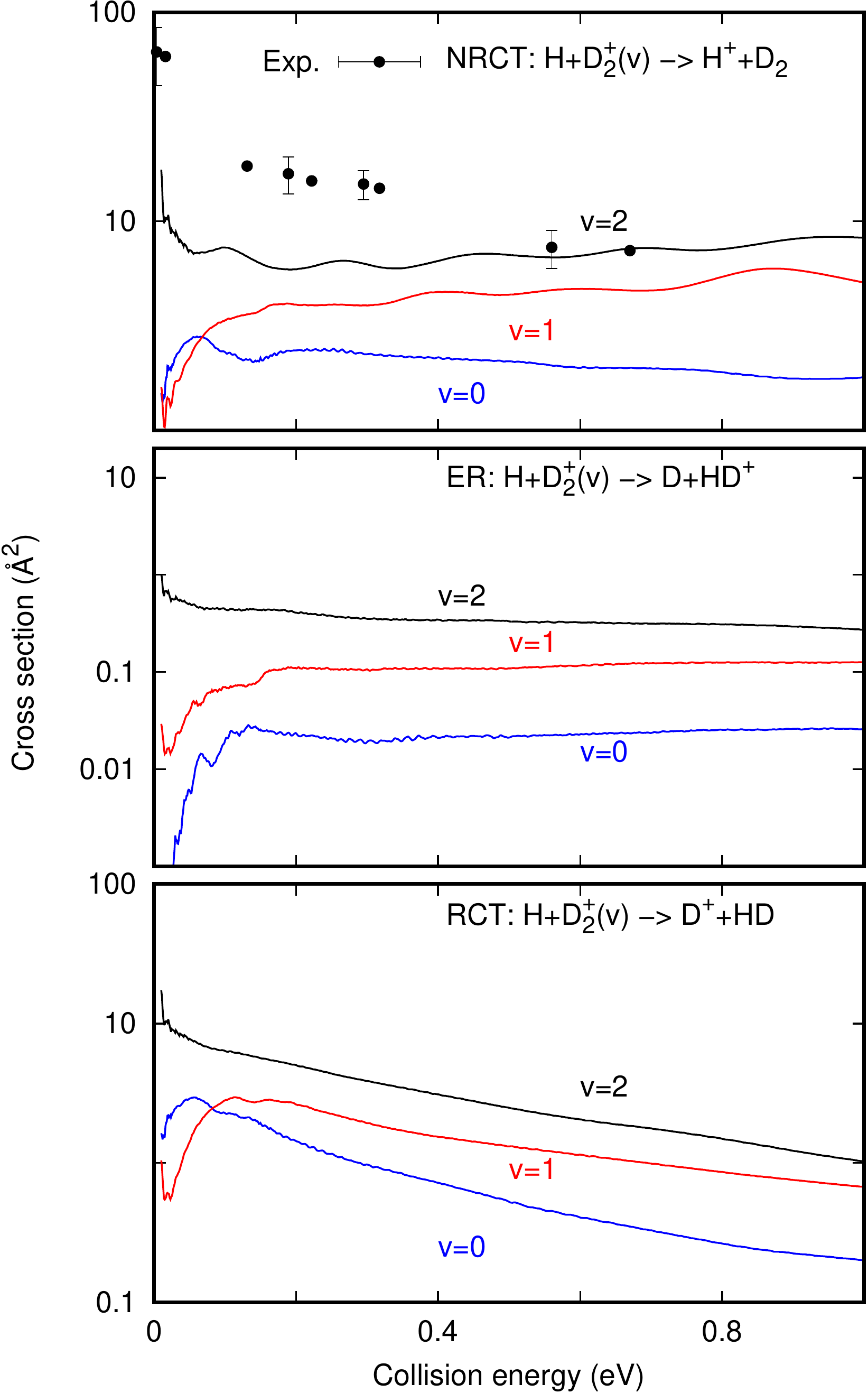}}}
\caption{Reaction cross section for  H+D$_2^+$(v=0, 1 and 2) 
    towards the inelastic charge transfer, NRCT (H$^+$+D$_2$, in the top panel),
    the exchange reactive channel, ER (H+HD$^+$, in the middle panel) and the
    reactive charge transfer, RCT, channels
    (D$^+$ +HD, in the bottom panel).
The experimental values are taken from Ref.\cite{Andrianarijaona-etal:09,Andrianarijaona-etal:19}.
  } \label{sigmatot}
\end{figure}

The cross section for the NRCT, ER, and RCT channels are shown in Fig.~\ref{sigmatot}.
Clearly the ER channel is one order of magnitude lower than the other two. NRCT and RCT
are of the same order at low energies, but as collision energy increases, RCT decreases
while NRCT is either constant or increases for v=1 and 2. At higher energy, it is
expected that this difference increases, becoming dominant the NRCT channel.
It should be noted that in the case of H+H$_2^+$ the two charge transfer processes,
RCT and NRCT,
are indistinguishible \cite{Sanz-Sanz-etal:21}. 

 The present theoretical
  results for each initial v are compared with the experimental cross sections in the top
  panel of Fig.~\ref{sigmatot}.
 In the experimental work by Andrianarijaona {\it et al}\cite{Andrianarijaona-etal:09,Andrianarijaona-etal:19}
  on the H + D$_2^+$ collisions, the H$^+$ products were detected, {\it i.e.} they provide
  a direct evidence on the NRCT cross section.
  The theoretical NRCT cross section increases as the initial D$^+_2$(v)
  vibrational state increases, and, for v=2
  at collision energies around 0.5 eV, it matches the corresponding experimental measurements. This
  seems to indicate that measurements made at lower temperature could be due to more vibrationally 
  excited initial D$_2^+$ reactants. In the experiments, the D$_2^+$ were produced by a CEA/Grenoble all-permanent magnet
  Electron Cyclotron Resonance ion source (ECR) \cite{Andrianarijaona-etal:09,Andrianarijaona-etal:19},
  and their vibrational distributions depend on the ECR ion source parameters,
  which are not coupled with the ion beam energy. Unfortunately,
  these experiments did not include any direct nor indirect measurement
  of the vibrational distributions of these D$_2^+$ ions. It is expected, though,
  that slight changes in the ion source conditions affect the vibrational distributions
  of the ions as observed in the case of molecular ions, including D$_2^+$,
  produced by Duoplasmatron sources \cite{Andrianarijaona:02}.
  A more detailed experimental work needs to be done with controlled conditions
  on the vibrational excitation of the D$_2^+$ reactant ions,
  specially at low collision energies, before a factual conclusion could be drawn.
  The 3-D imaging technique similar to the one used by Urbain {\it et al.} \cite{Urbain-etal:13}
  was proven to be efficient to measure vibrational distributions of H$_2^+$
  issued from the primary reaction H$^+$ +H$_2$.
  In this technique, two position sensitive detectors detect
  the positions and time of flights of fragments,
  giving access to the kinetic energy release,
  which definitely contains information on the vibrational state of the molecular ions.
  Given the similarity between Urbain {\it et al.}.’s molecular ions
  and the ones in Andrianarijaona {\it et al},’s experiment \cite{Andrianarijaona-etal:09,Andrianarijaona-etal:19},
  the 3-D imagine technique would be an appropriate fit to provide data that will accomplish our goal.
  The theoretical simulations presented in this work are expected to provide a good guidance
  to interpret such measurements. Moreover, this should also be extended to the H$_2^+$ + H reaction
  due to its astrophysical relevance, especially in Early Universe models.

In all the mechanisms, the cross section increases
with the  vibrational excitation. This is explained, as in the
reaction probabilities, by the increase
of the overlap between the initial D$_2^+$(v) and dominant D$_2$(v')
(see Fig.~\ref{pes-figure}.a),
which acts as
a doorway for the three mechanisms. The amplitude  of the initial vibrational
D$_2^+$(v) wave function around the electronic curve crossing region increases, thus
favoring
the electronic transition.

\begin{figure}
\centering
{%
  {\includegraphics[width=0.75\linewidth]{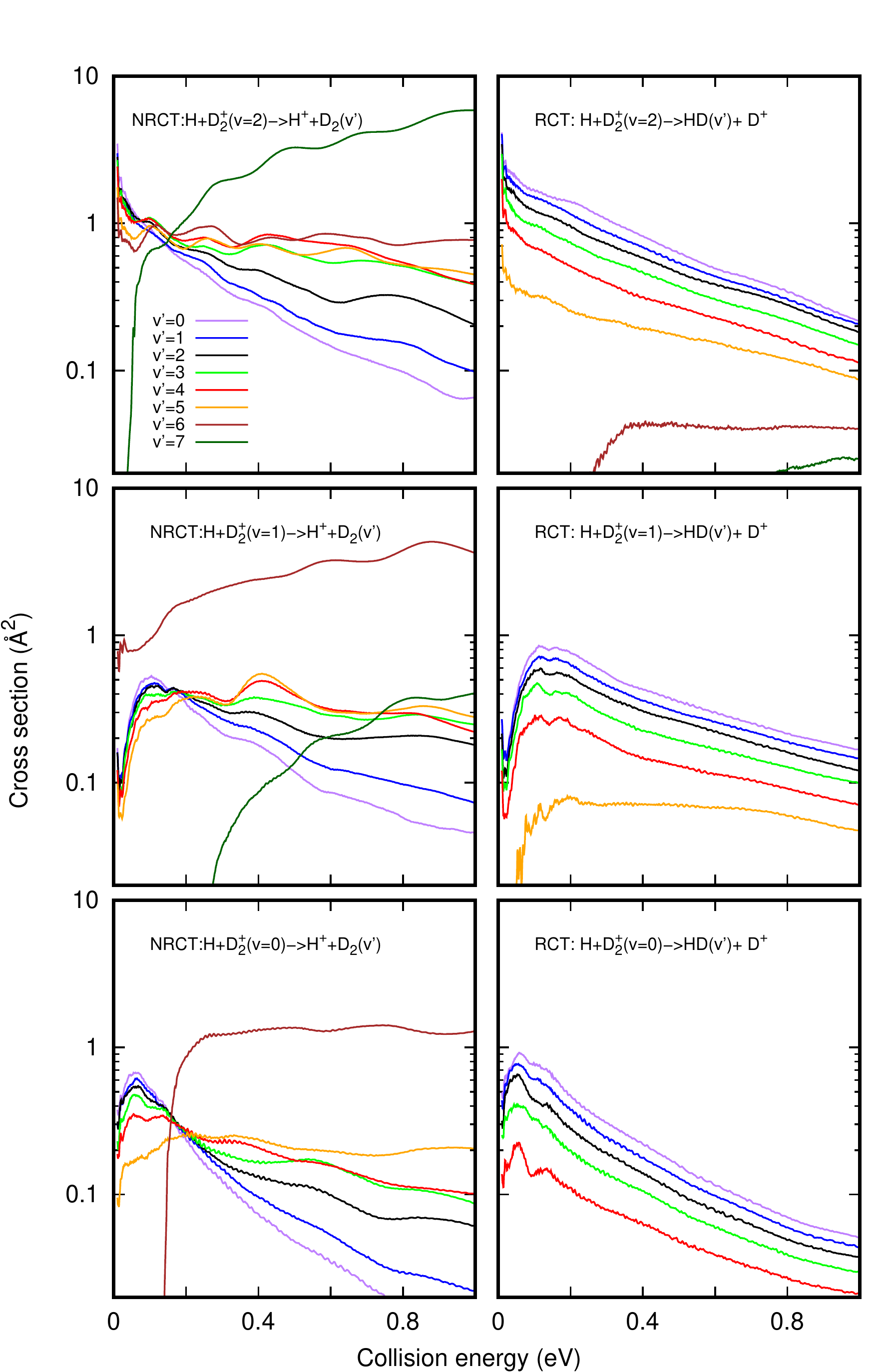}}}
\caption{Vibrationally resolved state-to-state cross section for  H+D$_2^+$(v=0, 1 and 2) 
    towards the inelastic charge transfer, NRCT (H$^+$+D$_2$(v'), in the left panels),
    and the reactive charge transfer, RCT, channels
    (D$^+$ +HD(v'), in the right panels).
  } \label{vibCTsigma}
\end{figure}

The vibrationally resolved state-to-state cross sections for the NRCT and RCT channels
are shown in Fig.~\ref{vibCTsigma}. There is a clear difference between the two mechanisms.
For RCT, all cross sections decrease with energy, which could be explained by the
near statistical mechanisms, due to the presence of many resonances associated
to the deep HD$_2^+$ well that mediate the reactivity.
For NRCT, however,  there are
several cases: there is always a dominant v': 
v'=6 and 7 for the cases of v=0,1 and v=2, respectively.  This
inelastic channel is formed by an electronic transition taking 
place at rather long distances. The rest of v' in the NRCT channel
show a progresively decreasing cross section with decreasing v', probably
due to transitions from v$_d$'.

\begin{figure}
\centering
{%
  \resizebox*{8cm}{!}
  {\includegraphics{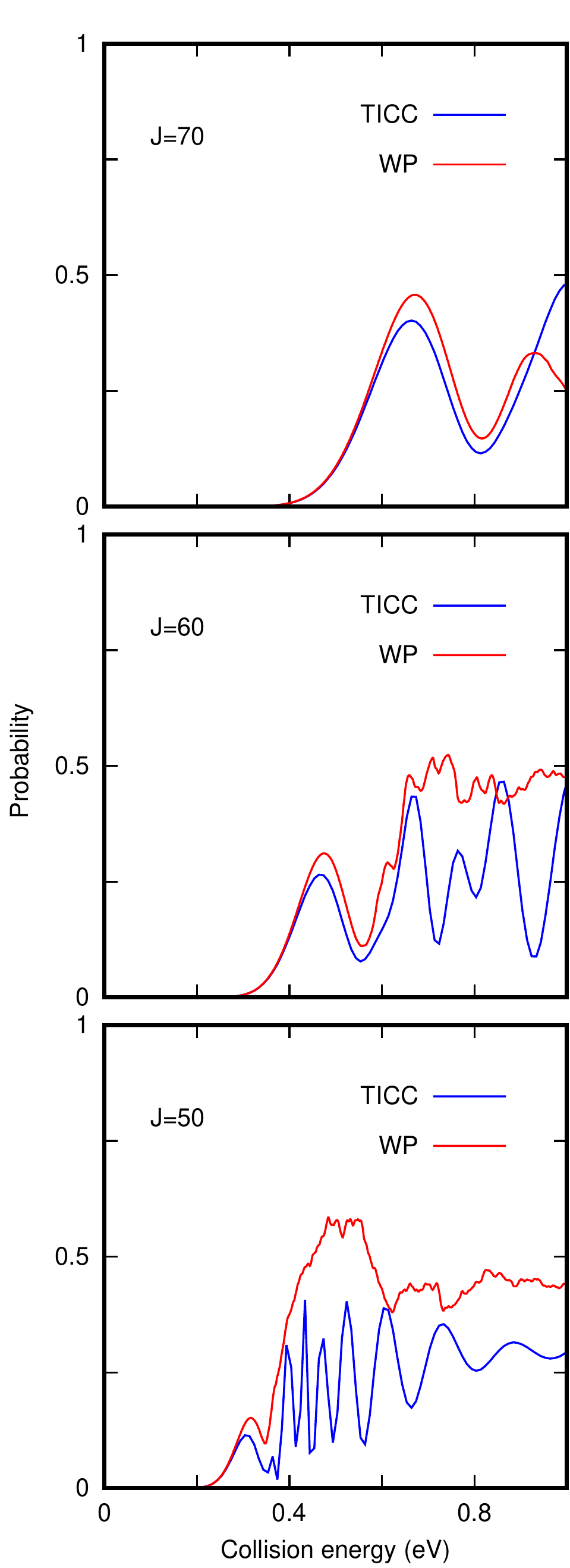}}}
\caption{NRCT probabilities obtained in H + D$_2^+$(v=2) collisions
  at several $J$ values using the ``exact'' wave packet
  results (WP) and the approximated time independent close coupling (TICC) results, as described in the text.
    } \label{NRCTticc}
\end{figure}

The reason why the NRCT channels  v'=6 and 7 for  D$_2^+$(v=0,1) and v'=7 for D$_2^+$(v=2)
are dominant at high energy can be easily interpreted in terms of the reaction
probabilities at high energies and high angular momentum. As $J$ increases,
the rotational barrier also increases. This barrier avoids the wave packet
to reach short distances between the two reactants, and therefore to enter
in the deep insertion well of the ground electronic state, where the reaction
takes place. However, the charge transfer between the initial D$_2^+$(v)
and nearly resonant D$_2$(v') can takes places at rather long distances,
at which small electronic couplings can induce transitions between close lying
electronic states. Under these circumstances, as reaction probabilities
decreases, the inelastic NRCT becomes dominant towards a particular
final D$_2$(v') state.

In order to check this model, we have performed approximate time-independent
close coupling (TICC) calculations, using reactant Jacobi coordinates,
including only one vibrational channel in the initial excited electronic
state, D$_2^+$(v=2), and one vibrational state in the final ground electronic
state,  D$_2$(v'=7), with 20 rotational channels in each, even $j$ values
from 0 to 40. The differential equations are integrated using the renormalized
Numerov method\cite{Gadea-etal:97,Roncero-etal:19} for distances
between 2 and 20 bohr, with 300 grid points. Also, the centrifugal sudden
approximation is assumed, keeping only one helicity, $\Omega$=0, since
the Coriolis coupling are expected to play a negligible role
at the long distances for which the electronic transition takes
place at  high total angular momentum.
  {
    These TICC calculations include the NRCT only, and moreover, only consider a possible CT vibrational
    product, D$_2$(v'=7). However, the WP presented above include all rearrangment channels, ER and RCT,
    and all accessible vibrational
    states of the products, becoming much more demanding computationally.
  }

The NRCT probabilities
obtained in these model calculations are compared in Fig.~\ref{NRCTticc}
with the accurate wave packet calculations described before for $J$=50, 60 and
70 (denotd by WP). Clearly, as $J$ increases, the agreement between the two methods improves.
This demonstrate the validity of the model. The oscillations obtained as a function
of collision energy are due to the interference between
the entrance and final channels energy difference\cite{Parker-etal:95,Tsien-Pack:70}.

This behavior obtained at high $J$ values shows that NRCT is 
the dominant reaction channel at high energy. This is valid not
only for the reaction studied in this work but also for all the other
isotopic variants, since it is attributed to the crossing between the
neutral and cationic diatomic fragments, which takes place at very
long distances between the reactants.

The TICC model has been extended to calculate the NRCT cross section in a wider
energy range, by solving the close-coupling equations for $J$= 0 to 300.
The simulated NRCT cross sections are compared with the experimental results
of Andrianarijaona {\it et al.}\cite{Andrianarijaona-etal:09,Andrianarijaona-etal:19}.
The TICC cross section is very close to the wave packet results for D$_2$(v'=7) below 1 eV,
but is lower than the total WP NRCT cross section, since it does not take into account
the other v' levels.
{
  As discussed above, the quasi degeneracy between initial D$_2^+$(v=2) and final
  D$_2$(v'=7) produces a very efficient electronic transition giving rise to the charge transfer.
  At low energy, for which relatively low total angular momentum $J$ are dominant, the rotational barriers
  are low and the dynamics in the ground electronic state can access to the deep well of H$_3^+$. Within this well
  there is an effective energy transfer towards the vibrational levels of D$_2$(v') and also access to HD products in
  other rearrangment channels. This explains why TICC results are so different to the accurate  WP results below 1eV, since
  TICC neglects these processes.
  However, at high angular momenta, which dominate at high collision energy,
  the rotational barrier does not allow the H and D$^+_2$ or H$^+$+D$_2$ reactants to approach each other
  at distances where the vibrational energy transfer or the exchange reaction can take place. 
}
This explains why 
the TICC approximation yields results in  rather good agreement with the experimental
results for collision energies between 3 and 9 eV.

\begin{figure}
\centering
{%
  \resizebox*{8cm}{!}
  {\includegraphics{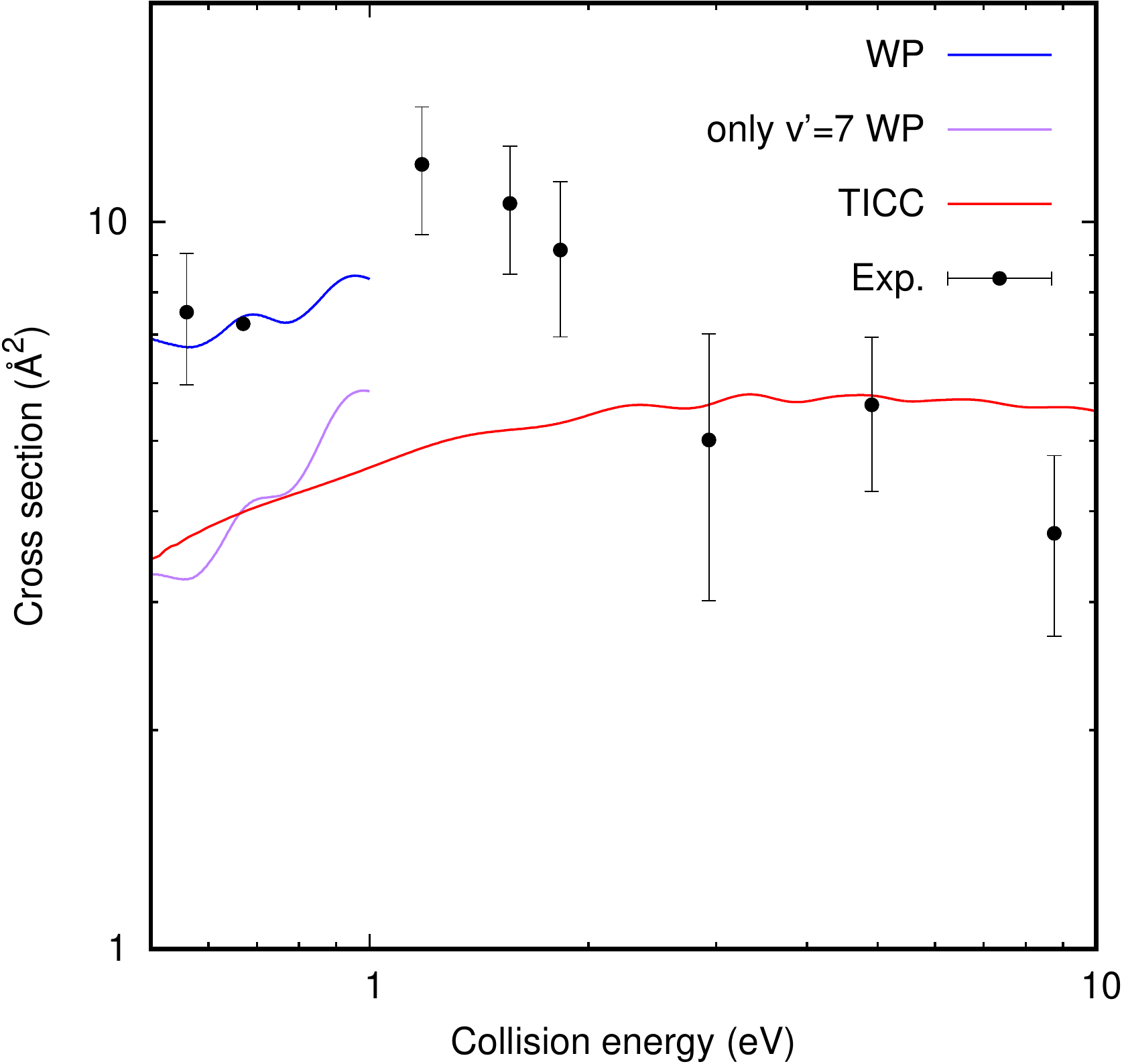}}}
\caption{NRCT cross section obtained in H + D$_2^+$(v=2) collisions
comparing the total NRCT  ``exact'' wave packet
results (WP), the state-to-state wave packet cross section for D$_2$(v'=7),
and the approximated time independent close coupling (TICC) results.
The experimental values are taken from Ref.\cite{Andrianarijaona-etal:09,Andrianarijaona-etal:19}.
    } \label{sigmaNRCTticc}
\end{figure}

\section{Conclusions}

The reactive cross sections for the H + D$_2^+$ charge transfer reaction increase
with the initial vibrational excitation of the D$_2^+$(v) reactant,
due to the increase of the amplitude of the vibrational
wavefunctions of D$_2^+$ and D$_2$ in the curve crossing and their mutual overlaps.
The dynamics is dominated by the electronic transition from the excited to
the ground electronic state, where the reaction takes place in
the deep insertion well of H$_3^+$. The charge transfer electronic
transition is dominated by the crossing of the two D$_2$/D$_2^+$ occurring
at rather large distance between reactants, where it can take place
between close lying vibrational states. For this reason, the reaction shows
a rather marked dependency on the initial vibrational excitation. Moreover,
this feature determines that 
the non-reactive charge transfer becomes the dominant channel as the collision energy increases,
for which the high total angular momentum introduces a barrier avoiding the access
to the insertion minimum.

The non-reactive charge transfer cross section is compared to the experimental
measurements in the 0.01-1 eV collision energy range studied theoretically.
A good agreement is found around 0.5 eV. At lower collision energies, the experimental
cross-section are larger than the experimental ones. This is attributted to a possible
higher vibrational excitation in the generation of D$_2^+$ reagents.
Also, a TICC model, including only the initial D$_2^+(v=2)$ and the near resonant D$_2$(v'=7) vibrational
states, shows good agreement with the experimental cross-section in the 3-9 eV collision energy range. 
It is concluded that
a more detailed experimental study, focusing on the control of experimental
conditions used to generate D$_2^+$, are needed to further analyze the vibrational
effects of reagents on the charge transfer reaction, specially at low energies. For
these studies, the theoretical simulations can be of important help.

\section*{Acknowledgement(s)}

We  acknowledge computing time at Cibeles (UAM)
under RES computational grant AECT-2021-1-0011.

\section*{Funding}

The research leading to these results has received fundings from
Ministerio de Ciencia, Investigaci\'on y Universidades (MICIU) (Spain) under grant FIS2017-83473-C2.
V. Andrianarijaona is supported by the
National Science Foundation through Grant No. PHY - 1530944


\end{document}